\documentclass[journal]{IEEEtran}
\usepackage{authblk}
\usepackage{diagbox}
\usepackage[compact]{titlesec}
\titlespacing{\section}{1pt}{*1}{*1}
\titlespacing{\subsection}{1pt}{*1}{*0}
\titlespacing{\subsubsection}{1pt}{*0}{*0}
\usepackage[
singlelinecheck=false 
]{caption}
\usepackage{graphicx}
\usepackage{mathtools}
\usepackage{comment}
\usepackage{algorithm,algpseudocode}
\algrenewcommand\algorithmicindent{0.5em}%
\usepackage{subfigure}
\usepackage{graphicx}
\usepackage{float}
\usepackage{amsfonts}
\usepackage{amsmath}
\usepackage{amsthm}
\usepackage{bm}
\usepackage{mathtools}
\usepackage{tabularx}
\usepackage{array}
\usepackage{breqn}
\usepackage{supertabular}
\usepackage{setspace}
\usepackage{upgreek}
\newcolumntype{H}{>{\setbox0=\hbox\bgroup}c<{\egroup}@{}}
\usepackage{url}
\usepackage{ragged2e}  

\newtheorem{lemma}{Lemma}
\newcolumntype{Y}{>{\RaggedRight\arraybackslash}X} 
\usepackage[export]{adjustbox}
\usepackage{xcolor}
\DeclareCaptionFont{blue}{\color{blue}}
\usepackage[inline]{trackchanges}
\addeditor{NG}

\begin{document}

\title{
Large-Scale Maintenance and Unit Commitment: \\ A Decentralized Subgradient Approach
}


\author{\IEEEauthorblockN{Paritosh Ramanan\IEEEauthorrefmark{1}\IEEEauthorrefmark{2},Murat Yildirim\IEEEauthorrefmark{3}, Nagi Gebraeel\IEEEauthorrefmark{2} and Edmond Chow\IEEEauthorrefmark{1}}
\thanks{\IEEEauthorrefmark{1}School of Computational Science and Engineering, Georgia Institute of technology, Atlanta, GA, USA 30332}
\thanks{\IEEEauthorrefmark{3}College of Engineering, Wayne State University, Detroit, MI, USA 48202}
\thanks{\IEEEauthorrefmark{2}H. Milton Stewart School of Industrial and Systems Engineering, Georgia Institute of technology, Atlanta, GA, USA 30332.\\paritoshpr@gatech.edu,murat@wayne.edu,nagi.gebraeel@isye.gatech.edu,\\echow@cc.gatech.edu\\
This research is based upon work supported by the U.S. Department of Energy, Office of Fossil Energy (National Energy Technology Laboratory), under Award Number DOE-FE0031288.}
}
\maketitle
\begin{abstract}
Unit Commitment (UC) is a fundamental problem in power system operations. When coupled with generation maintenance, the joint optimization problem poses significant computational challenges due to coupling constraints linking maintenance and UC decisions. Obviously, these challenges grow with the size of the network. With the introduction of sensors for monitoring generator health and condition-based maintenance(CBM), these challenges have been magnified. ADMM-based decentralized methods have shown promise in solving large-scale UC problems, especially in vertically integrated power systems. However, in their current form, these methods fail to deliver similar computational performance and scalability when considering the joint UC and CBM problem. 

This paper provides a novel decentralized optimization framework for solving large-scale, joint UC and CBM problems. Our approach relies on the novel use of the subgradient method to temporally decouple various subproblems of the ADMM-based formulation of the joint problem along the maintenance horizon. By effectively utilizing multithreading, our decentralized subgradient approach delivers superior computational performance and eliminates the need to move sensor data thereby alleviating privacy and security concerns. Using experiments on large scale test cases, we show that our framework can provide a speedup of upto 50x as compared to various state of the art benchmarks without compromising on solution quality. 
\end{abstract}
\begin{IEEEkeywords}
Sensor driven prognosis, Joint operations and condition based maintenance, Decentralized and multithreaded optimization, Vertically integrated power systems.
\end{IEEEkeywords}
\section*{Nomenclature}
\vspace{-2mm}
\textbf{Sets}:
\vspace{-2mm}
\begin{center}
\begin{supertabular}{c lc}
$\mathcal{N}_r,G_{r},\mathcal{U}_{r},\mathcal{V}_{r},\mathcal{I}_{r}$ & Neighboring regions, generators,  foreign,\\& boundary \& internal buses of region $r$.\\[1mm]
$G^b_{r},\mathcal{U}^b_r,\mathcal{V}^b_{r},\mathcal{I}^b_r$ & Generators, boundary, foreign \& internal \\& buses connected to bus $b \in \mathcal{U}_{r} \cup \mathcal{I}_r$\\[1mm]
$\mathcal{R}$ & The set of all regions\\[1mm]
$\mathcal{B}_r$ & $\mathcal{U}_r \cup \mathcal{V}_{r}$, Boundary, foreign buses of $r$.\\[1mm]
$\mathcal{N}^b_r$ & Neighboring regions connected to bus \\& $b \in \mathcal{U}_{r}$\\[1mm]
$\mathcal{B}^b_r$ & $\mathcal{U}^b_r \cup \mathcal{V}^b_{r} \cup \mathcal{I}^b_r$, Neighboring buses of $b$. \\[1mm]
$M$ & Maintenance planning horizon \\[1mm]
$T$ & Operational planning horizon \\
\end{supertabular}
\end{center}
\vspace{5mm}
\textbf{Decision Variables} (at $t\in T$):
\begin{center}
\begin{supertabular}{c lc}
$y^{g}_t$ & The electricity dispatch of generator $g$\\[1mm]
$x^{g}_t \in \{0,1\}$ & The commitment decision variable of $g$\\[1mm]
$z^{g}_t \in \{0,1\}$ & The maintenance variable of generator $g$\\[1mm]
{$\psi^{b}_{t}$} & Demand Curtailment at bus $b$\\[1mm]
$\theta^{b}_{t}$ & The phase angle at bus $b$\\[1mm]
$\tilde{\theta}^{b,r' }_t$ & The phase angle of bus $b$
where $b \in \mathcal{U}_{r'}$ \\ & and $r' \in \mathcal{N}_r$\\[1mm]
$\pi^g_{Ut}, \pi^g_{Dt} $ & The up and down variable of generator g\\[1mm]
$p_{r,t}$ & Production difference at region $r$\\[1mm]
$f^{uv}_{t}$ & Power flow from bus $u$ to $v$ such that $u \in \mathcal{U}_{r}$ \\& and $v \in \mathcal{V}^u_{r}$\\[1mm]
  {$\alpha_g$} &   {The Lagrangian multiplier for maintenance} \\&   {cardinality constraint of generator g}\\[1mm]
$\lambda^{b}_t$ & The Lagrangian multiplier with respect to \\&phase angles of bus b where $b \in \mathcal{U}_{r} \bigcup \mathcal{V}_{r}$\\[1mm]
$\phi^{uv}_t$ & The Lagrangian multiplier with respect to flow\\& from bus $u$ to bus $v$ where $u \in \mathcal{U}_{r}$ and $v\in \mathcal{V}_{r}$\\& for any region $r$\\[1mm] 
$\eta_{r,t}$ & The Lagrangian multiplier with respect to \\&production target at region $r$\\[1mm]
\end{supertabular}
\end{center}

\textbf{Constants}:
\begin{center}
\begin{supertabular}{c lc}
$d^{g},c^{g},S^g_U,S^g_D$ & The dispatch, commitment , start-up and \\& shut-down cost of generator $g$\\[1mm]
$\mu^g_U, \mu^g_D, R^g$ & Minimum up time, down time and ramp-up \\& ramp-down constant for $g$\\[1mm]
$\delta^{b}_t$ & The demand at bus $b$ at $t \in T$\\[1mm]
$F^{uv}_{max}$ & Maximum capacity of line  connecting  buses \\& $u$ and $v$  such that $u \in \mathcal{U}_{r}$ and $v \in \mathcal{V}^u_{r}$\\[1mm]
$\rho_{\theta},\rho_{f}$ & Penalty parameter for phase angles, flows \\[1mm]
$\Gamma^{uv}$ & Phase angle conversion for line $uv$\\[1mm]
\end{supertabular}
\end{center}

\section{Introduction}\label{sec:intro}
Power network operations has traditionally been represented by the well-studied UC problem \cite{unit_commitment}. The UC problem optimizes generator commitment and production decisions subject to network topology, transmission, and generation constraints. Maintenance of generation assets like generators and turbines has received a fair share of attention in the literature \cite{maintDereg, maintDereg2}. 
However, most of the research has focused on periodic and calendar-based maintenance schedules where the goal is to optimize maintenance intervals to minimize maintenance cost \cite{maintDereg2}.  Few research efforts have tried to integrate maintenance with network operations. Some noteworthy examples include \cite{marwali1998integrated, fu2007security,fu2009coordination}.   {These problems typically focus on a vertically integrated power system setting involving periodic maintenance schemes. Today, the rise of digital frameworks like the Internet-of-Things (IoT) is transforming the power generation industry. It has also highlighted the importance of leveraging IoT technologies like condition monitoring to advance maintenance management, i.e., condition-based maintenance (CBM). As a result, recent works have studied the integration of condition based maintenance (CBM) with operations and demonstrated significant cost savings by incorporating sensor-based predictive analytics \cite{muratp1, muratp2}.} This paper provides a novel decentralized optimization framework to address the computational challenges in solving the joint UC and CBM optimization problem for large scale power networks. 

The joint UC and CBM problem is significantly more complex than the standard UC problem because it involves maintenance and commitment coupling constraints. Moreover, the planning horizon for the joint problem tends to be in the order of months due to the timeline of maintenance decisions, whereas for UC it is typically 24 hours. Given the long planning horizons, the   {number of} binary variables related to maintenance decisions are orders of magnitude greater than the conventional UC problem setting. 

From a computational standpoint, decentralized solution methodologies, such as alternating direction method of multipliers (ADMM), have been widely used to improve computational performance of large scale mixed integer planning problems like UC  \cite{javad,ramanan2017asynchronous,asynch2019uc,xavier2020decomposable}. These solution methodologies rely on decomposing the power network topology into multiple autonomous regions. The methods iteratively apply ADMM by dualizing network flow constraints corresponding to transmission lines between regions. However, conventional ADMM-based decentralization approaches are not sufficient to solve large scale instances of the joint UC and CBM problem. This is because the local regional subproblems themselves become orders of magnitude more complex than the corresponding regional subproblems found in decentralized UC. Therefore, a direct adaptation of ADMM-based decentralization is not sufficient to address the computational complexities of the joint problem. 

  {
In addition to computational benefits, an added capability of decentralized methods also rests in its ability to significantly reduce or eliminate the need to move data \cite{javad,asynch2019uc,xavier2020decomposable}. The digitization of the grid due to an influx of IoT enabled sensors, has led to important questions on the security, privacy and handling of sensor data. In fact, the U.S. Department of Homeland Security's Industrial Control Systems (ICS) Cyber Emergency Response Team indicates that the energy sector accounts for nearly 35\% of all ICS-related incidents in the United States\cite{doe_plan}. Moreover, within power systems, estimating operating conditions of generation equipment followed by communication between multiple subsystems have been deemed essential system functions likely at risk to cyber attacks \cite{osti_1337873}.
Needless to say, the reliance of CBM methodologies on IoT driven sensor data makes it a perfect candidate for cyber attacks and potential misuse of sensor datasets by malicious parties. The need to frequently transmit highly resolved sensor data from a vast number of assets to a centralized database amplifies the risks of potential data leaks by providing a greater attack surface for intruders \cite{osti_1337873}. As a result, data manipulation attacks could take place which could lead to inaccurate estimates of asset health \cite{li2020degradation,li2020detection} directly affecting maintenance policies and the overall stability of the grid. By preserving sensor data privacy and eliminating the need for its transmission, decentralized methods significantly reduce the risks of data leaks leading to enhanced security measures.
}

While decentralization has several benefits, the unique aspects of our joint maintenance and UC problem present numerous computational complications. The basis of our approach rests on the efficient decomposition of large scale instance of the joint UC and CBM optimization problem along the temporal and regional domains. Our approach relies on the novel use of the subgradient method to temporally decouple various regional subproblems of the ADMM-based problem formulation along the maintenance time horizon. The regional decomposition is solved in a decentralized fashion while the temporal decomposition is leveraged for multithreading. 

Instead of directly solving each subproblem locally as proposed in \cite{javad},\cite{asynch2019uc},\cite{async_opt} the introduction of the subgradient method enables us to deftly leverage multithreading to transform the complex joint problem formulation into a highly scalable, computationally efficient one. Next, we iteratively employ ADMM \cite{admm_boyd} to balance power flow between neighboring regions by dualizing the flow constraints corresponding to respective tie lines. Our specific contributions in this paper are summarized as follows: 


\begin{itemize}

\item We develop a computationally efficient, decentralized, multithreaded subgradient method for solving large instances of the joint UC and CBM optimization problem. Existing ADMM-based decentralization models \cite{async_opt}, \cite{asynch2019uc}, alone are not sufficient to solve such large scale instances. Therefore, our method relies on exploiting the block diagonal structure obtained by dualizing the maintenance cardinality constraints. This paves the way for applying the subgradient method to the local subproblems which can effectively leverage multithreading. Our method therefore, transforms the joint formulation into a massively parallelizeable solution framework.


\item
Using our decentralized approach, we eliminate the need to move any critical sensor data required for maintenance scheduling at every region. This respects data residency requirements and significantly reduces cybersecurity threats that can arise from leaked IoT data. 


\item For evaluation purposes, we develop a High Performance Computing (HPC) implementation of our decentralized framework based on a hybrid Message Passing Interface (MPI) and OpenMP frameworks. We assign every region to an MPI process and multiple OpenMP threads to every region to achieve a scalable and efficient implementation. 
\end{itemize}

The complexity of the joint problem is evidenced by the fact that the solution to the IEEE 3012 bus case over a planning horizon of 3 months required approximately 48 hours by state-of-the-art commercial solvers. On the other hand, our multithreaded, decentralized solution paradigm delivered a speedup of up to 50 in some cases by effectively leveraging the computational power of 1200 threads orchestrated on a state-of-the-art supercomputer. The nature and scale of compute employed in this paper is in stark contrast to existing large-scale, decentralized UC implementations \cite{asynch2019uc,xavier2020decomposable,async_opt} which only required using up to 120 threads for the same problem instances.

In Section \ref{sec:mrw}, we review relevant literature pertaining to decentralized optimization of power systems. Details of the decentralized CBM problem formulation are discussed in Section \ref{sec:mpf}. In Section \ref{sec:malg}, we develop the decentralized joint CBM and operations algorithm that preserves regional data privacy. In Section \ref{sec:mresults}, we present our experimental results followed by conclusion and future work in Section \ref{sec:mconclusion}.

\section{Related work}\label{sec:mrw}
Large scale problems in power systems are characterized by long planning horizons and large network sizes. While distributed and decentralized algorithms have both been applied towards solving large scale power system problems, there is a subtle but profound difference between the two categories. Distributed algorithms like those based on cutting planes are typically characterized by the presence of a master process that coordinates the entire computational progress \cite{ramanan2017asynchronous}. On the other hand, decentralized algorithms, while being parallel and distributed in nature, function even in the absence of a master process \cite{asynch2019uc}.
As a result, decentralized approaches yield two important benefits that are not possible otherwise in generic distributed approaches relying on a master process. Due to the presence only of peer-to-peer message exchanges, decentralized approaches exhibit greater scalability \cite{xavier2020decomposable}. Further, unlike their distributed counterparts, decentralized methods are based on a topology decomposition of the power network into multiple regions that are representative of utilities or their subsidiaries thereof \cite{javad}. Each region retains full ownership of local data and only exchanges flow information pertaining to shared transmission lines with neighbors \cite{javad,asynch2019uc}. 

Another important differentiator between generic distributed algorithms and decentralized approaches is concerning the implementation aspect. As a consequence of their reliance only on peer-to-peer message exchanges, a decentralized algorithm can be implemented on a geographically distributed computing architecture \cite{ramanan2017asynchronous, asynch2019uc}. On the other hand, decompositions in distributed algorithms are designed mainly for computational convenience with the intent of unleashing massive parallelization potential \cite{hpcuc,uc_async}. Typically, subproblems in distributed algorithms emanate from temporal decompositions for deterministic optimization cases \cite{muratp2} and scenario driven decompositions for stochastic cases \cite{hpcuc,uc_async} without representing any real world entity. Therefore, such approaches can only function on a specialized computational system like an HPC cluster that offers significant parallelization potential but cannot handle geographically distributed problem instances \cite{ramanan2017asynchronous}. Owing to such tight limitation of the implementation aspect, distributed approaches force end users to transfer relevant information to an HPC cluster leading to data privacy concerns \cite{xavier2020decomposable}. 

Distributed solutions pertaining to operations have conventionally revolved around stochastic UC \cite{hpcuc,uc_async} with scenario based decompositions.  The work done in \cite{hpcuc} solves stochastic unit commitment problem for a fleet of sustainable energy sources where the imposed demand is time varying. The authors in \cite{uc_async} solve a unit commitment problem using an incremental subgradient method to progress the dual variable while simultaneously recovering primal feasibility. Lately, decentralized approaches for solving deterministic UC are gaining popularity for their enhanced scalability and privacy preserving aspects as stated above \cite{javad,ramanan2017asynchronous,asynch2019uc,xavier2020decomposable}.

On the other hand, as mentioned in Section \ref{sec:intro}, efforts that have focused on integrating maintenance and network operations are limited and based on the cutting plane method. Pertinent examples include \cite{marwali1998integrated, fu2007security,fu2009coordination} where models for optimizing networks operations integrated periodic maintenance of assets and transmission lines. In  \cite{abiri2012optimized}, the authors integrated reliability information, namely generator failure rates, in a simplified operations problem. The authors in \cite{muratp1,muratp2} have proposed a sensor data driven, joint optimization of CBM with operations which represents the current state of the art. However, this approach does not scale well with larger instances on account of limited potential for parallelization. Therefore, a decentralized technique to solve the joint problem would be highly beneficial in order to achieve scalability as well as reduce computational time significantly. However, decentralized solution paradigms specifically tailored for the joint maintenance problem have largely remained unexplored.

\section{Problem Formulation}\label{sec:mpf}
\begin{figure}[!htb]
\centering
\includegraphics[width=0.48\textwidth,keepaspectratio]{ 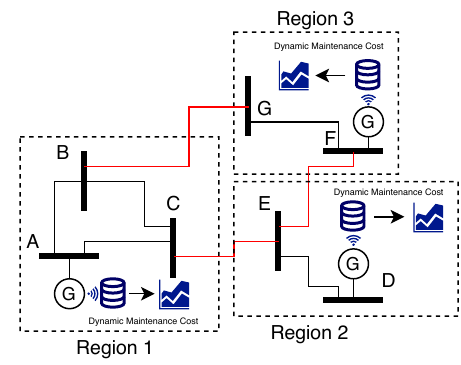}
\caption{Partition of Network topology into regions with full data privacy.}
\label{fig:sample_synchronous} 
\end{figure}

We propose a decentralized formulation based on regional decomposition with multiple regional subproblems. From a practical standpoint, each region may denote a subsidiary of the utility company in a vertically integrated market. Consider the sample network in Figure 
\ref{fig:sample_synchronous} consisting of 3 regions with boundary and foreign bus categorization defined as follows:
\begin{itemize}
\item Region 1: $\mathcal{U}_{1} = \{B,C\}$, $\mathcal{V}_{1} = \{G,E\}$
\item Region 2: $\mathcal{U}_{2} = \{E\}$, $\mathcal{V}_{2} = \{F,C\}$
\item Region 3: $\mathcal{U}_{3} = \{G,F\}$, $\mathcal{V}_{3} = \{B,E\}$
\end{itemize}
Every region is comprised of local (critical) generators and buses subject to its own operational constraints. Every critical generator is instrumented with sensors for monitoring health/degradation. Sensor data in each region is streamed to respective local databases where predictive analytic algorithms are used to predict the remaining operational life of the generator. As mentioned earlier, the focus of this paper is not to develop accurate predictive degradation models but rather to address the computational challenges in large scale joint optimization of operations and maintenance in the presence of IoT-enabled generation assets. 
\subsection{Degradation-based Predictive Modeling}
Predictive analytics is a crucial aspect of our approach. Degradation-based sensor data from IoT-enabled generators are incorporated into the regional operations and maintenance scheduling problems. We leverage contemporary degradation models developed by \cite{ bayesdegrad} to compute predictions of generator remaining lifetimes. The authors in \cite{ bayesdegrad} proposed a Bayesian framework that utilizes real-time degradation signals from partially degraded assets to predict and continuously update residual life distributions (RLDs). This framework was later extended in \cite{elwany2008sensor} by computing optimal replacement/maintenance and spare parts ordering policies driven by real-time RLD predictions. The authors used the RLDs to calculate a convex maintenance cost function to identify the optimal time to perform maintenance. The cost function represents the trade-off between the cost associated with the risk of unexpected failure and the opportunity cost associated with performing premature (or unnecessary) preventive maintenance. As new sensor data is observed, an asset's RLD and the corresponding cost function are updated through a Bayesian framework. \cite{muratp1,muratp2} adopted this approach in their joint optimization problem and successfully integrated these dynamically evolving expected maintenance cost functions. Considering a maintenance planning horizon $M$, the expected maintenance cost function at maintenance interval $m \in M$ is expressed in {Equation \ref{eq:mcostfunc}.
\begin{equation}\label{eq:mcostfunc}
    \omega^{g}_{m} = \kappa\cdot\frac{ \omega_p^g P( \tau_g > m) + \omega_f^g P( \tau_g \leq m) }{ \int_0^{m} P( \tau_g > z) dz + m_o }
\end{equation}}
In Equation \ref{eq:mcostfunc}, $\omega^{g}_{m}$ denotes the maintenance cost at epoch $m$ for generator $g$ of age $m_o$. $\kappa$ represents the maintenance criticality coefficient, denoting the relative importance of maintenance with respect to operations. $\omega^g_p$ and $\omega^g_f$ are the costs for preventive maintenance, and unexpected failure for generator $g$, respectively while $\tau_g$ is the remaining life of generator $g$. In other words, the function $\omega^{g}_{m}$ translates $\tau_g$ (the RLD of generator $g$) into a degradation-based maintenance cost function over time.
\subsection{Decentralized Joint Maintenance and Operations}
We note that the maintenance planning horizon set $M$ comprises of multiple smaller operational planning horizons $T_m$ such that,
\[T =\bigcup\limits_{m=1}^{|M|}T_m,\text{ where, }T_m = \Bigg\{t | t\in \Bigg[\frac{m|T|}{|M|}\ldots\frac{(m+1)|T|}{|M|}\Bigg]\Bigg\}  \]
Therefore, the regional subproblem seeks to minimize the objective cost as represented by Problem \eqref{eq:OPT_OBJ}. 
\begin{equation}\label{eq:OPT_OBJ}
\begin{aligned}
\mathcal{L}_r(\Delta)=&\sum_{t\in T}\sum_{g\in  G_r}\Big[c^g_{t}x^g_{t}+d^g_{t}y^g_{t}+S^g_U\pi^g_{Ut}        +S^g_D\pi^g_{Dt}\Big]\\
            +&\sum_{t\in T}\Big[E_t(\bm{\bar{\theta}},\bm{\phi}) + H_t(\bm{\bar{F}}, \bm{\lambda}) + I_t(\bm{\bar{p}},\bm{\upeta})\Big]\\
            +&\nu\sum_{t\in T}\sum_{u \in \mathcal{U}_{r} \cup \mathcal{I}_r}\hspace{-3mm}\psi^u_t + \sum_{g\in  G_r}\sum_{m \in M} \omega^g_mz^g_m
\end{aligned}
\end{equation}
where $\Delta = \{\bm{\bar{\theta}},\bm{\bar{F}},\bm{\bar{p}},\bm{\lambda}, \bm{\phi}, \bm{\upeta}\}$ represents the consensus phase angle, flow, regional production target and their respective lagrangian multipliers. Further,
\[E_t(\bm{\bar{\theta}}, \bm{\lambda})=\sum\limits_{b \in \mathcal{B}_{r}}	\Big[\lambda^{b}_{t} (\theta^b_t - \bar{\theta}^b_{t})+\frac{\rho_\theta}{2} (\theta^b_t - \bar{\theta}^b_{t})^2 \Big]\]
\[H_t(\bm{\bar{F}},\bm{\phi}) =\sum\limits_{u \in \mathcal{U}_{r}}\sum\limits_{v \in \mathcal{V}^u_{r}}	\Big[\phi^{uv}_{t}(F^{uv}_t -\bar{F}^{uv}_{t,k})+ \frac{\rho_f}{2} (F^{uv}_t - \bar{F}^{uv}_{t,k})^2 \Big] 
\]
\[I_t(\bm{\bar{p}},\bm{\upeta}) = \upeta_{r,t}(p_{r,t}-\bar{p}_{r,t}) + \frac{\rho_p}{2}(p_{r,t}-\bar{p}_{r,t})^2\]

The objective function represented by Problem \eqref{eq:OPT_OBJ}, consists of a commitment, operations and maintenance cost components. In addition, it also consists of ADMM penalty terms imposed on balancing flow estimates among neighboring regions. Flow estimates are iteratively balanced across transmission lines common with neighboring regions as well as with respect to the phase angles at their respective boundary buses. We attempt to reduce overall demand violation by determining a local, customized production target $\bar{p}_{r,t}$ based on global demand violation. Computing a customized production target for every region has been shown to yield smooth convergence in large scale decentralized methods especially from the operational standpoint \cite{asynch2019uc}. 

In our decentralized formulation represented by Problem \eqref{eq:OPT_OBJ}, the operational decision including commitment and production are decoupled { at linking intervals for every maintenance epoch \cite{fu2007security,fu2009coordination,muratp1,muratp2,wu2010security}. 
The decoupling of minimum up/down, ramping constraints at linking intervals is a common practice meant to decompose the computationally challenging joint problem into multiple short-term UC subproblems as illustrated in \cite{fu2007security,fu2009coordination,wu2010security}. 
}
Therefore, for a particular maintenance epoch $m\in M$, let $\bm{Q}^r_m$ denote the set of inequalities \eqref{eq:op_mt_cpl}-\eqref{eq:sq8}, $\forall t\in T_m$.
\begin{subequations}\label{eq:OPT}
\allowdisplaybreaks[1]
\begin{align}
&x^g_t \leq 1- z^g_m, \quad \forall g \in G_r \label{eq:op_mt_cpl}\\[1mm]
&P^g_{min}x^{g}_t \leq y^g_t \leq P^g_{max}x^{g}_t,
\quad \forall g \in G_r & \label{eq:sq1}\\[1mm]
&-\pi^g_{Dt} \leq x^g_t -x^g_{t-1} \leq \pi^g_{Ut}, 
\ \forall g \in G_r & \label{eq:sq2}\\[1mm]
&-R^g \leq y^g_t - y^g_{t-1} \leq R^g,  
\ \forall g \in G_r & \label{eq:sq3}\\[1mm]
\begin{split} & \ \Gamma^{uv}(\theta^u_t - \theta^v_t) = f^{uv}_t, \ \forall u \in \mathcal{U}_{r}, \forall v \in \mathcal{V}^u_{r}\\
\end{split} & \label{eq:sq4}\\[1mm]
\begin{split} & -F^{uv}_{max}\leq \Gamma^{uv}(\theta^u_t - \theta^v_t) \leq F^{uv}_{max}, 
\ \forall u \in \mathcal{U}_{r} \cup \mathcal{I}_r,\forall v \in \mathcal{B}^u_{r}\end{split} & \label{eq:sq5}\\[1mm]
\begin{split} & \sum\limits_{\forall g \in G^u_{r}}y^g_t - \delta^{u}_t +\psi_t^u = \sum\limits_{\forall v \in \mathcal{B}^u_{r}}[\Gamma^{uv}(\theta^{u}_t - \theta^{v}_t)],\forall u \in \mathcal{U}_{r} \cup \mathcal{I}_r\end{split}\label{eq:sq7}\\[1mm]
\begin{split} & \sum\limits_{\forall i\in U_t} \pi^g_{Ui} \leq x^g_t \leq 1-\sum\limits_{\forall i \in D_t} \pi^g_{Di}, \ \forall g \in G_r, 
\\ & \qquad\qquad\ U_t=[t-\mu^g_U+1,t], D_t = [t-\mu^g_D+1,t] \end{split} \label{eq:sq8}
\end{align}
\end{subequations}
Constraint \eqref{eq:op_mt_cpl} ensures that a generator that has been placed under maintenance must not have any production. Constraint \eqref{eq:sq1} enforces production limits at each generator while Constraints \eqref{eq:sq2} and \eqref{eq:sq8} enforce minimum up and down-time for each generator. Constraint \eqref{eq:sq3} enforces their respective ramping limitations. Equation \eqref{eq:sq4} establishes the linear relationship between flow and their respective phase angles. Constraint \eqref{eq:sq5} enforces transmission line capacity constraints. Equation \eqref{eq:sq7} balances the demand at each bus with local generation and network flow. Equations \eqref{eq:sq4}-\eqref{eq:sq7} enforce network flow constraints globally.
Therefore, the joint CBM and operations problem can be represented by Problem \eqref{eq:OPT_aug}.
\begin{subequations}\label{eq:OPT_aug}
\allowdisplaybreaks[1]
\begin{align}
{\text{min}} & \ \mathcal{L}_r(\Delta) & & \\
\text{s.t.} & \ \sum_{m\in M} z^g_m = 1, \quad \forall g \in  G_r \label{eq:card}\\[1mm]
&\bm{x},\bm{y},\bm{z},\bm{\pi}_U,\bm{\pi}_D,\bm{\theta},\bm{f}\in\bm{Q}^r_m,\quad \forall m \in M \label{eq:qi}
\end{align}
\end{subequations}
In Problem \eqref{eq:OPT_aug} Constraint \eqref{eq:card} represents the maintenance cardinality constraint. Specifically, Constraint \eqref{eq:card} ensures that maintenance must be performed on each generator exactly once during the maintenance planning horizon. { We also note that our formulation represented by Problem \eqref{eq:OPT_aug} does not make any assumptions on the length of the maintenance epochs. Therefore, the same decentralized, subgradient based approach proposed in this section is capable of handling different values of lengths of maintenance epochs (like 2 weeks, 3 weeks etc.).}

We estimate two important consensus quantities, i.e. \emph{intermediate flow} denoted by $\bar{F}^{uv}_t$ and \emph{intermediate phase angles} denoted by $\bar{\theta}^{b}_t$ based on Equations \eqref{eq:tintm},\eqref{eq:fintm} respectively. These intermediate values are calculated based on the phase angle estimates received from neighboring regions $ \forall b \in \mathcal{B}_r$ and $\forall u \in \mathcal{U}_{r}, \forall v \in \mathcal{V}_{r}$ for phase angles and flows respectively applied $\forall t \in T$.
\begin{equation}\label{eq:tintm}
\bar{\theta}^{b}_t = \frac{\theta^b_t + \sum\limits_{r' \in \mathcal{N}^b_r} \tilde{\theta}^{b,r' }_t}{|\mathcal{N}^b_r|+1}, \quad b\in \mathcal{B}^r
\end{equation}
\begin{multline}\label{eq:fintm}
\bar{F}^{uv}_t = \frac{\Gamma^{uv}(\tilde{\theta}^{u,r'}_t - \tilde{\theta}^{v,r'}_{t}) + \Gamma^{uv}(\theta^{u}_t - \theta^{v}_{t})}{2}, \\  u \in  \mathcal{U}_{r}, v\in \mathcal{V}^u_{r}, r' \in \mathcal{N}^u_{r}
\end{multline}
We update the Lagrangian multipliers $\lambda,\phi$ based on Equations \eqref{eq:lupdt},\eqref{eq:fupdt} respectively.
\begin{equation}\label{eq:lupdt}
\lambda^{b}_t = \lambda^{b}_t+ \rho_\theta(\theta^b_t - \bar{\theta}^b_t), \quad  \forall b \in \mathcal{B}_r, \forall t \in T
\end{equation}
\begin{equation}\label{eq:fupdt}
\phi^{uv}_t = \phi^{uv}_t + \rho_f(F^{uv}_t - \bar{F}^{uv}_t), \quad \forall u \in \mathcal{U}_{r}, \forall v \in \mathcal{V}_{r}, \forall t \in T
\end{equation}
We estimate the customized production target for region $r$ based on Equation \eqref{eq:dv}.
\begin{equation}\label{eq:dv}
    \bar{p}_{r,t} =  \sum\limits_{\forall g \in G_r} y_t^g + \frac{\sum\limits_{\forall r \in \mathcal{R}} \varphi_{r,t} }{|\mathcal{R}|}
\end{equation}
In Equation \eqref{eq:dv}, $\varphi_{r,t} = \Big|\sum\limits_{\substack{\forall u \in \mathcal{U}_r\bigcup\mathcal{I}_r}}\hspace{-4mm}(\delta^u_t-\psi^u_t) - \sum\limits_{\substack{\forall g \in G_r}} y_t^g\Big|$ represents the local violation $\forall t \in T$. We update the associated Lagrangian $\upeta$ based on Equation \eqref{eq:pupdt}.
\begin{equation}\label{eq:pupdt}
    \upeta_{r,t} = p_{r,t} + \rho_p(p_{r,t}-\bar{p}_{r,t})
\end{equation}
Therefore, the joint optimization model given by Problem \eqref{eq:OPT_aug} describes a Mixed-Integer Quadratic Problem (MIQP) which solves for the maintenance and operations in a decentralized manner. 
The Lagrangian terms in the model serve as \emph{penalties} for deviating from a position of balance. Convergence occurs when the dualized flow terms become small enough so that the optimization problem given by (\ref{eq:OPT_aug}) become mathematically equivalent to that of a \emph{centralized} problem as described in \cite{muratp2}. 
We are now in a position to state Lemma \ref{thm1}.
\begin{lemma}\label{thm1}
Let $\mathcal{L}_r(\Delta)$ represent the objective of Problem \eqref{eq:OPT_aug} subject to constraints \eqref{eq:card},\eqref{eq:qi}, then
\[\sum\limits_{m\in M}\underset{\bm{Q}^r_{m}}{\text{min }} \mathcal{L}^m_r(\Delta,\bm{\alpha}) \leq {\text{min }}\mathcal{L}_r(\Delta)\]
where, $\alpha_g$ represents the dual variable with respect to Constraint \eqref{eq:card}
\end{lemma}
\begin{proof}
Dualizing Equation \eqref{eq:card} with the dual variable $\alpha_g$, we obtain the Lagrangian relaxation $\hat{\mathcal{L}}_r$
\begin{equation}\label{eq:LR}
    \hat{\mathcal{L}}_r = \mathcal{L}_r(\Delta)+\sum\limits_{g\in  G_r}\alpha_g(1-\sum\limits_{m\in M}z^g_m) =  \sum\limits_{m\in M} \mathcal{L}^m_r(\Delta,\bm{\alpha})
\end{equation}
where,
\begin{subequations}
\begin{align*}
\mathcal{L}^m_r(\Delta,\bm{\alpha})&=\hspace{-2mm}\sum_{t\in T_m}\Big[E_t(\bm{\bar{\theta}},\bm{\phi}) + H_t(\bm{\bar{F}}, \bm{\lambda})+ I_t(\bm{\bar{p}},\bm{\upeta}) + \nu\hspace{-4mm}\sum_{u \in \mathcal{U}_{r} \cup \mathcal{I}_r}\hspace{-3mm}\psi^u_t \Big] \\
&\hspace{-15mm}+\sum_{g\in  G_r}\Bigg[\sum_{t\in T_m}\Big[c^g_{t}x^g_{t}+d^g_{t}y^g_{t}+S^g_U\pi^g_{Ut}        +S^g_D\pi^g_{Dt}\Big]+(\omega^g_{m})z^g_{m} \Bigg]  \\
&\hspace{-15mm}+\sum\limits_{g\in  G_r}\alpha_g\left(\frac{1}{|M|}-z^g_m\right)
\end{align*}
\end{subequations}
Let $\bm{\bar{\vartheta}} = \{\bm{x},\bm{y},\bm{z},\bm{\pi}_U,\bm{\pi}_D,\bm{\theta},\bm{f}\}$ be the optimal solution to $\mathcal{L}_r(\Delta)$ (i.e. Problem \eqref{eq:OPT_aug}). With this solution fixed, $\bm{\bar{z}}$ ensures $
(1-\sum\limits_{m\in M}z^g_m) = 0 \quad \forall {g\in  G_r}$, which would enforce i) $\sum\limits_{g\in  G_r}\alpha_g(1-\sum\limits_{m\in M}z^g_m) = 0$, and ii)  $\hat{\mathcal{L}}_r |_{ \bm{\bar{\vartheta}}} = \mathcal{L}_r(\Delta)|_{ \bm{\bar{\vartheta}}}$. 

Let $\bm{{\vartheta}}'$ and $\mathcal{F}$ be the optimal solution, and the set of feasible solutions for $\hat{\mathcal{L}}_r$, respectively. Note that the solution ${ \bm{\bar{\vartheta}}}$ is still feasible for the lagrangian relaxation (i.e. ${ \bm{\bar{\vartheta}}} \in \mathcal{F}$), it is trivial to conclude $\hat{\mathcal{L}}_r |_{ \bm{{\vartheta}}'} = \min_{\vartheta \in \mathcal{F}} \hat{\mathcal{L}}_r |_{ \bm{{\vartheta}}} \leq \hat{\mathcal{L}}_r |_{ \bm{\bar{\vartheta}}}$. Since, $\hat{\mathcal{L}}_r |_{ \bm{{\vartheta}}'} \leq \hat{\mathcal{L}}_r |_{ \bm{\bar{\vartheta}}} = \mathcal{L}_r(\Delta)|_{ \bm{\bar{\vartheta}}}$, we conclude:
\begin{equation}
    \underset{\bigcup\limits_{m\in M}\bm{Q}^r_{\bm{m}}}{\text{min }}\hat{\mathcal{L}}_r =\sum\limits_{m\in M}\underset{\bm{Q}^r_{\bm{m}}}{\text{min }} \mathcal{L}^m_r(\Delta,\bm{\alpha})  \leq {\text{min }}\mathcal{L}_r(\Delta)
\end{equation}
We note that minimizing $\hat{\mathcal{L}}_r$ is equivalent to minimizing each of the individual terms $\mathcal{L}^m_r$ since their respective constraint sets $\bm{Q}^r_{\bm{m}}$'s are disjoint $\forall m\in M$.
\end{proof}
As a result of Lemma \ref{thm1}, in order to perform one solve of Problem \eqref{eq:OPT_aug} it suffices to iteratively solve $\mathcal{L}^m_r$ in a multithreaded parallel fashion $\forall m\in M$ followed by an update of $\alpha_g$ using the subgradient method outlined in \cite{fisher}.
\section{Algorithm Design for Joint Decentralized Maintenance and Operations}\label{sec:malg}
We divide the joint, decentralized optimization algorithm into four distinct parts. Each component pertains to the local multithreaded solver, peer-to-peer communication scheme, decentralized optimization frameworks based on convex relaxation and the subgradient methods respectively. 
\subsection{Local Multithreaded Optimization Solver}
\begin{algorithmic}\label{alg:lmtopt}
\Function{MTOpt}{$\alpha,\mathcal{L}_r(\Delta),\bm{Q}^r$}
    \For{$m=1,2\ldots|M|\text{ \textbf{using multithreading}}$}
    \State solve $\underset{Q^r_m}{\text{min }} \mathcal{L}^m_r(\Delta,\alpha)$ using Lemma \ref{thm1}
    \EndFor
    \State \Return $\{\bm{x},\bm{y},\bm{z},\bm{\theta},\bm{f},\bm{\varphi}\}$
\EndFunction
\end{algorithmic}
The entire decentralized joint problem relies on a local optimization solver represented by function $\texttt{MTOpt}$. Owing to its completely decoupled nature along the maintenance planning horizon as detailed in Lemma \ref{thm1}, we apply multithreading to solve Problem \eqref{eq:OPT_aug} in parallel to boost computational efficiency. 
Therefore, given an objective function $\mathcal{L}_r(\Delta)$, a Lagrangian estimate $\alpha$ of Constraint \eqref{eq:card} and a constraint set $\bm{Q}^r$, function $\texttt{MTOpt}$ applies $M$ threads to solve one iteration of the local joint problem.

\subsection{Peer to peer Communication}
\begin{algorithmic}
\Function{Communicate}{$k,\Delta^{k-1},\bm{\theta}^k,\bm{f}^k,\bm{\varphi}^k$}
\State send $\bm{\theta}^{b,k}$ to all regions $r',$ $\forall r' \in \mathcal{N}_r, b\in \mathcal{B}_{r'}$
        \State recieve $\bm{\tilde{\theta}}^{b,k}$ from all regions $r',$ $\forall r' \in \mathcal{N}_r, b\in \mathcal{B}_{r'}$
        \State send $\bm{\varphi}^k$ and recieve $\bm{\varphi}^k_{r'}$ to and from all regions $r' \in \mathcal{R}$
        \State compute $\Delta^{k}$ based on Equations \eqref{eq:tintm}-\eqref{eq:pupdt}
        \If {$||\bm{f}^{k}-\bm{\bar{f}}^k||<\epsilon$ and $||\bm{\bar{f}}^{k}-\bm{\bar{f}}^{k-1}||<\epsilon$}
        	\State set local convergence to true
        	\State if local convergence is true $\forall r \in \mathcal{R}$ then $\Upomega\leftarrow 1$
	    \EndIf
	    \State \Return $\{\Delta^{k},\Upomega\}$
\EndFunction
\end{algorithmic} 
Communication between neighboring regions occurs according to the steps represented in function \texttt{Communicate} which is invoked after every local solve. 
At every round $k$, estimates of the phase $\bm{\theta}^k$, flow $\bm{f}^k$, local violation $\bm{\varphi}^k$, their corresponding consensus values embodied in $\Delta^{k-1}$ are taken as inputs. The phase angle estimates are sent to neighbors and are used to compute fresh estimates of the intermediate values $\Delta^{k}$. The local load violation $\bm{\varphi}^{r,k}$ is also communicated to all regions and is used to compute the customized local production target. A local convergence check follows based on the consensus estimates of the flow.   {The global convergence status is denoted by $\Upomega$. In case of local convergence of all regions, the problem is deemed to have globally converged and $\Upomega$ is set to 1.}
\subsection{Decentralized Optimization with Fixed Maintenance}
\begin{algorithmic}
\Function{DecentFixedOpt}{$\Delta,\bm{Q}^r$}
\State $k\leftarrow 0$, $\Delta_0\leftarrow \Delta$, $\alpha\leftarrow 0$
\State set global convergence value $\Upomega\leftarrow0$
\While {$\Upomega$!$=1$}
\State $k\leftarrow k+1$
\State $\{\bm{x}^{k},\bm{y}^{k},\bm{z}^{k},\bm{\theta}^k,\bm{f}^k,\bm{\varphi}^k\}$$\leftarrow$$\Call{MTOpt}{\alpha,\mathcal{L}_r(\Delta^{k-1}),\bm{Q}^r}$
\State $\{\Delta^{k},\Upomega\}\leftarrow\Call{Communicate}{k,\Delta^{k-1},\bm{\theta}^k,\bm{f}^k,\bm{\varphi}^k}$
\EndWhile
\State \Return $\{\bm{x}^k,\bm{y}^{k},\bm{z}^{k},\Delta^k\}$
\EndFunction
\end{algorithmic}
For solving the joint problem with the maintenance decisions fixed, we follow the sequence of steps represented in function \texttt{DecentFixedOpt}. Since maintenance decisions are fixed, the Lagrangian $\alpha$ is also fixed to 0. A multithreaded local solve followed by a peer to peer communication forms one round of the decentralized joint framework. The sequence of computation followed by communication occurs synchronously among all regions until global convergence is achieved.

\subsection{Decentralized Optimization with Local Subgradients}
\begin{algorithmic}
\Function{DecentSGOpt}{$\Delta,\bm{Q}^r$}
\State $k\leftarrow 0$, $\Delta_0\leftarrow \Delta$, $\alpha^0\leftarrow 0$
\State set global convergence value $\Upomega\leftarrow0$
\While {$\Upomega$!$=1$}
\State $k\leftarrow k+1$, $j\leftarrow 0$, $\Delta^{sg}\leftarrow\Delta^{k-1}$
\While{Constraint \eqref{eq:card} not satisfied}
\State $\{\bm{x}^{j},\bm{y}^{j},\bm{z}^{j},\bm{\theta}^j,\bm{f}^j,\bm{\varphi}^j\}\leftarrow\Call{MTOpt}{\alpha^j,\mathcal{L}_r(\Delta^{sg}),\bm{Q}^r}$
\State $\mathcal{L}^j=\bm{c}_r^T\bm{x}^j+\bm{d}_r^T\bm{y}^j+\bm{\omega}_r^T\bm{z}^j$ 
\State   {$\alpha^{j+1}_g = \alpha^{j}_g + \sigma^j(1-\sum\limits_{m \in M}z^{g,j}_{m})$, $\forall g \in G_r$}
\State update $\sigma^{j+1}$ based on Equation \eqref{eq:sgupdt}
\State $\{\bm{x}^{k},\bm{y}^{k},\bm{z}^{k},\bm{\theta}^k,\bm{f}^k,\bm{\varphi}^k\}$$\leftarrow$$\{\bm{x}^{j},\bm{y}^{j},\bm{z}^{j},\bm{\theta}^j,\bm{f}^j,\bm{\varphi}^j\}$
\EndWhile
\State $\{\Delta^{k},\Upomega\}\leftarrow\Call{Communicate}{k,\Delta^{k-1},\bm{\theta}^k,\bm{f}^k,\bm{\varphi}^k}$
\EndWhile
\State \Return $\{\bm{x}^k,\bm{y}^{k},\bm{z}^{k},\Delta^k\}$
\EndFunction
\end{algorithmic}
When the maintenance decisions are released, we utilize the sequence of steps detailed in function \texttt{DecentSGOpt}. Specifically, the framework outlined in \cite{fisher} is utilized to update the Lagrangian value $\alpha_g$   { which is iteratively updated using the Equation \ref{eq:alpha_updt}
\begin{equation}\label{eq:alpha_updt}
    \alpha^{j+1}_g = \alpha^{j}_g + \sigma^j(1-\sum\limits_{m \in M}z^{g,j}_{m}), \text{  ,}\forall g \in G_r
\end{equation}}
The step size for the subgradient method is given by Equation \eqref{eq:sgupdt} where $\mathcal{L}^j,\mathcal{L}_{UB}$ represent the joint operational and maintenance objective costs of the $j^{th}$ local subgradient iteration and its upper bound respectively.
\begin{equation}\label{eq:sgupdt}
    \sigma_j = \frac{|\mathcal{L}_{UB}-\mathcal{L}^j|}{\sum\limits_{g \in  G_r}\Big[1-\sum\limits_{m \in M}z^{g,j}_{m}\Big]^2 }
\end{equation}
\subsection{Decentralized Multithreaded Joint Maintenance Algorithm}
\begin{algorithm}
\caption{Decentralized Multithreaded(DMT) Algorithm for Joint CBM and Operations}\label{alg:syncd}
\begin{algorithmic}
\State \hspace{-4mm} $\{\bm{x},\bm{y},\bm{z},\Delta\}_{FMRC}$$\leftarrow$$ \Call{DecentFixedOPT}{\Delta_0,\bm{Q}^r_{FMRC}}$
\State \hspace{-4mm} $\{\bm{x},\bm{y},\bm{z},\Delta\}_{FMBC}$$\leftarrow$$ \Call{DecentFixedOPT}{\Delta_{FMRC},\bm{Q}^r_{FMBC}}$
\State \hspace{-4mm} compute $\mathcal{L}_{UB}$ based on $\{\bm{x},\bm{y},\bm{z}\}_{FMBC}$
\State \hspace{-4mm} $\{\bm{x},\bm{y},\bm{z},\Delta\}$$\leftarrow$$ \Call{DecentSGOpt}{\Delta_{FMBC},\bm{Q}^r_{BMBC}}$
\end{algorithmic}
\end{algorithm}
\begin{figure}
    \centering
    \includegraphics[width=0.49\textwidth,keepaspectratio]{ 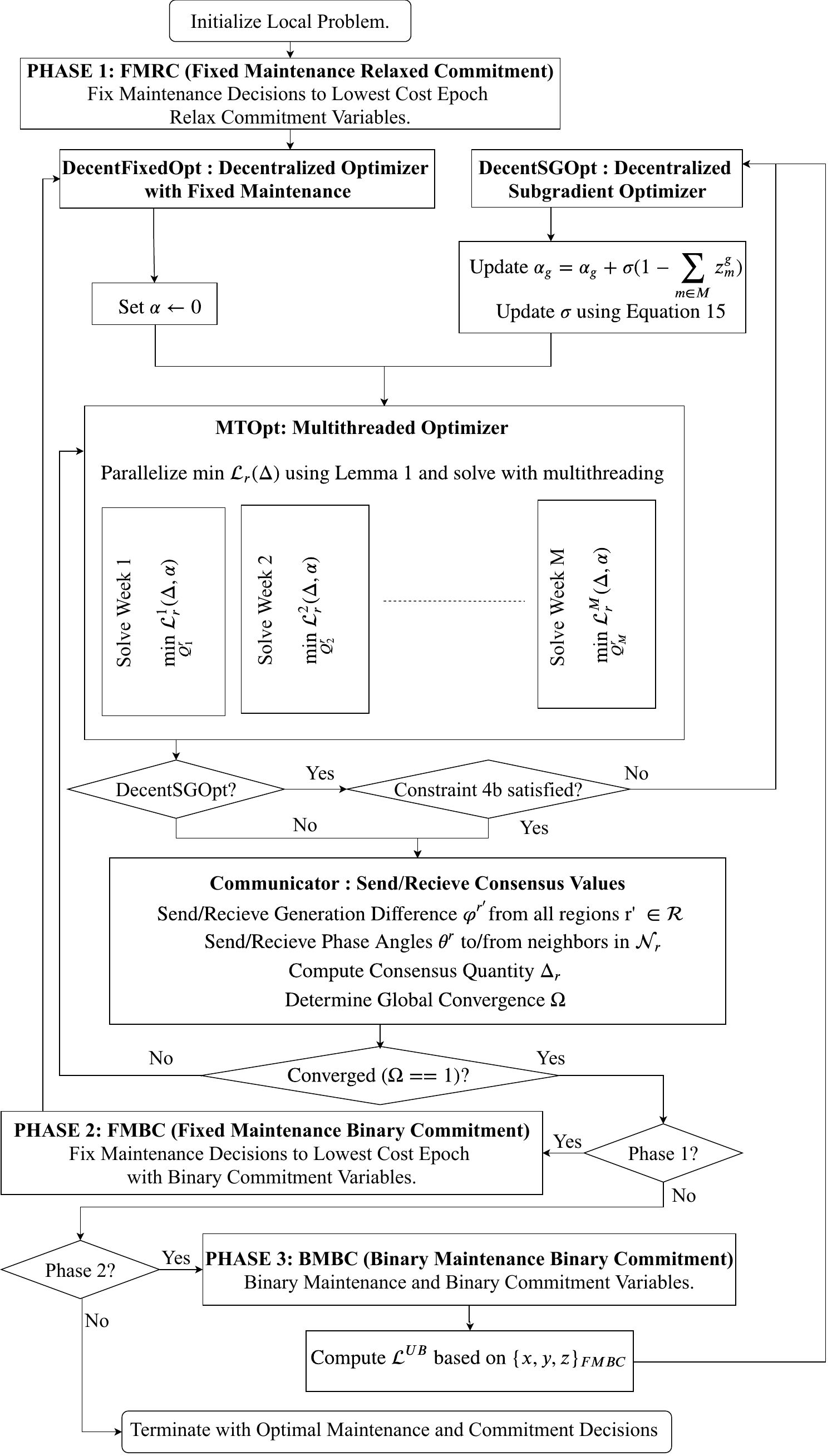}
    \caption{Flowchart depicting the DMT algorithm at region $r$}
    \label{fig:flowchart}
\vspace{-5mm}
\end{figure}
\begin{figure*}
    \centering
    \includegraphics[width=\textwidth,keepaspectratio]{ 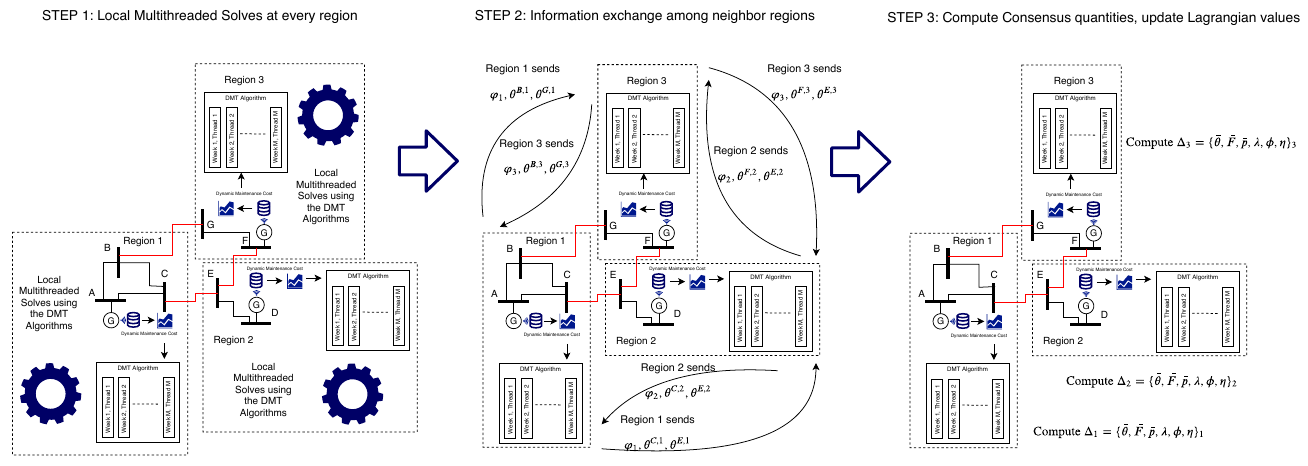}
    \caption{Schematic describing the DMT algorithm in the context of peer-to-peer consensus of phase angle and flow values}
    \label{fig:overall_schematic}
\vspace{-3mm}
\end{figure*}
Based on the four main functions mentioned before, we now present the decentralized multithreaded (DMT) joint optimization solution framework in Algorithm \ref{alg:syncd}. We begin by initializing the intermediate values and delineating the three different constraint sets $\bm{Q}^r_{FMRC},\bm{Q}^r_{FMBC}$ and $\bm{Q}^r_{BMBC}$. Set $\bm{Q}^r_{FMRC}$ consists of maintenance decisions fixed to the lowest epoch of the maintenance cost function with relaxed commitment variables. As a result, $\bm{Q}^r_{FMRC}$ is convex in nature. Further, $\bm{Q}^r_{FMBC}$ represents the constraint set with fixed maintenance and binary commitment variables. It is easy to see that an optimal solution of Problem \eqref{eq:OPT_aug} subject to $\bm{Q}^r_{FMBC}$ is a feasible solution and also an upper bound of the joint problem as well. Moreover, a solution to Problem \eqref{eq:OPT_aug} with respect to $\bm{Q}^r_{FMBC}$ is purely operations oriented. Finally, $\bm{Q}^r_{BMBC}$ represents a constraint set with released maintenance variables and binary commitment variables.

Initially, Algorithm \ref{alg:syncd} globally converges with respect to $\bm{Q}^r_{FMRC}$. Next, the resulting intermediate values are used as input to converge with respect to $\bm{Q}^r_{FMBC}$. Based on the optimal solution obtained in the previous step, an upper bound of the joint solution is determined that in turn forms the basis of the subgradient method. Warm starting with the help of the intermediate values of the previous step, the local solves based on the subgradient method globally converge to yield the joint maintenance and operations solution. 

  {
A flowchart of the DMT algorithm has been provided in Figure \ref{fig:flowchart}. The flowchart reflects the program logic of Algorithm \ref{alg:syncd} at a regional level. The process starts with an initialized local problem, where the maintenance variables would be fixed, and relaxed commitment decisions would be optimized using the Phase 1 (Fixed Maintenance Relaxed Commitment) formulation. 
The balance of consensus quantities achieved among neighbor regions in Phase 1 is used to warm start Phase 2 (Fixed Maintenance Binary Commitment), wherein the binary decisions are obtained with respect to the maintenance decisions fixed in Phase 1. Upon converging with respect to Phase 2, an upper bound on the joint CBM and operations problem is obtained regionally. Next, the subgradient method is instantiated across all regions which in turn takes advantage of the consensus achieved thus far. The subgradient method in Phase 3 is applied iteratively until Constraint \eqref{eq:card} is locally met to complete one regional iteration. These regional iterations are performed by all regions until global convergence is achieved.}

  {
While Figure \ref{fig:flowchart} presents the regional program flow, Figure \ref{fig:overall_schematic} depicts the ADMM based coordination employed in Algorithm \ref{alg:syncd} based on the sample power network illustrated in Figure \ref{fig:sample_synchronous}. Broadly, there are three main steps concerning the consensus process. Under Step 1, the regions perform multithreaded, local computation using the three different phases illustrated in Figure \ref{fig:flowchart}. After local computation, in Step 2, the phase angles and generation difference values are exchanged with neighbors and all other regions respectively. In Step 3, the received values from other regions are utilized to compute the respective $\Delta$ values which are used as an input in the subsequent iteration of DMT.
}
\section{Experimental Results}\label{sec:mresults}
For our experiments, we used \verb|Gurobi 7.5| \cite{gurobi} for solving the MIQP problem represented by (\ref{eq:OPT}) on each node. All our experiments were run with 128GB of memory on a state of the art super computer \cite{PACE}. We evaluate our model on the IEEE 3012 bus case with data derived from \verb|MATPOWER| library \cite{matpower}. In order to model the regions, we utilize the region decompositions as well as the data provided in \cite{javad} for the IEEE 3012 bus case for 150 generators. {In all our experiments, the penalty terms $\rho_{\theta}, \rho_f$ and $\rho_p$ are fixed at a value of 1.}

\subsection{Computational Architecture}\label{sec:comparch}
\begin{figure}[!htb]
    \centering
    \includegraphics[width=0.48\textwidth,keepaspectratio]{ 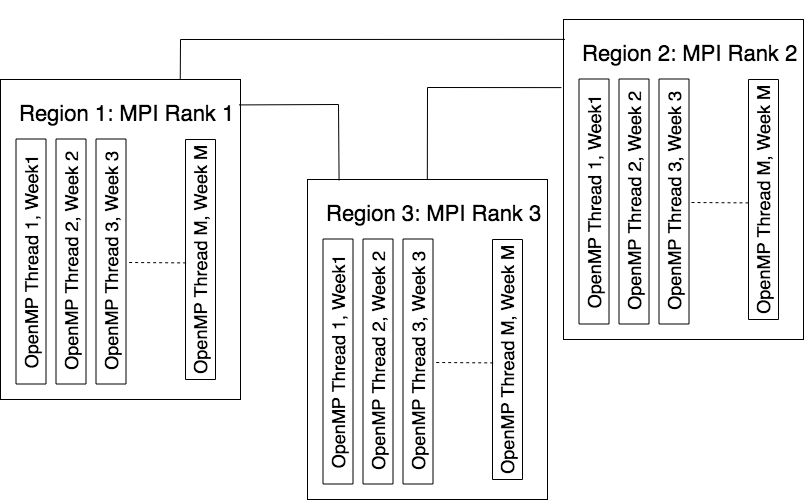}
    \caption{Hybrid MPI OpenMP Computational Architecture for solving the Joint CBM and Operations problem}
    \label{fig:comparch}
\end{figure}

For our experiments, we use an implementation that is based on the Message Passing Interface (MPI)\cite{mpi4py} framework for decentralized computation and OpenMP for multithreading. The computational architecture for the sample power network depicted in Figure \ref{fig:sample_synchronous} is presented in Figure \ref{fig:comparch}. The architecture follows a hybrid MPI-OpenMP architecture where each region is assigned one MPI process and every maintenance epoch for the regional subproblem is assigned to one OpenMP thread. Using a distributed memory framework like MPI for decentralization helps us evaluate the algorithm in an environment close to the \emph{real-world}, where each region may represent various utilities. Our multithreaded decentralized computational architecture and software framework can be easily extended to a geographically distributed computational environment. 
\subsection{Degradation Modeling}
In order to obtain the sensor data necessary to derive the maintenance cost of the assets, we rely on vibration data acquired from a rotating machinery apparatus. Using this experimental setup, condition monitoring is employed to estimate the degradation of the rolling element bearing present in the rotating machine apparatus. We chose roller bearings because they represent an integral component of every rotating machinery including generators and different types of turbines \cite{cbmepp,rollingbearing}. Vibration signals from the bearings are used to represent generator degradation. We follow an experimental setup outlined in \cite{bayesdegrad} that traces the degradation of bearings from their new state until their failure.

\subsection{Benchmark Models}
  {For all our experiments, we consider a planning horizon of 12 weeks. To highlight the benefits of our decentralized multithreaded approach, we resolve the problem under different operational complexity cases. More specifically, the operational complexity in our setting is quantified by the commitment decisions per generator per day (CGD).
In order to evaluate the computational benefits of our framework, we consider four different benchmark models.
\begin{enumerate}
    \item \textit{Centralized Benchmark}: Our centralized benchmark comprises a direct solve of the large scale problem formulation by Gurobi without any decomposition schemes. We use the centralized benchmark to rate the performance of the decentralized algorithm by measuring the relative optimality gap. We let $\gamma$ denote the total optimal objective value comprised of operations and CBM components. $\gamma_{decent} \text{ and } \gamma_{c}$ represent the optimal objective for the decentralized joint method and its centralized counterpart, respectively.
    Solving a centralized version of the joint problem could be significantly challenging in itself. In our experiments, the centralized version did not converge within 6 hours for even the simplest operational complexity of 4 CGD. Therefore, our centralized benchmark formulation considers its relaxation without transmission line constraints. By relaxing both the integrality constraints of the commitment decisions, and the transmission line constraints in the centralized benchmark, we report the performance of the other models as a function of a conservative lower bound for the centralized problem. This means that the optimality gaps reported by the decentralized solution may actually be lower.   {We also note that an exact theoretical rate of convergence cannot be provided owing to the mixed integer and     consequently NP-hard nature of the joint problem as indicated in \cite{async_opt}.  We do show that our method provides a solution quality that consistently provides an optimality gap of approximately 1\%.}
    \item \textit{Two Stage APMII Benchmark}: The Two Stage APMII Benchmark is based on a modified Benders decomposition strategy introduced in \cite{muratp2}. The joint formulation is decomposed into a maintenance-driven master problem as well as weekly operational subproblems that are captured through Benders and integer-type cuts. 
    Driven inherently by a Benders decomposition, this two stage method represents the current state of the art for solving the joint CBM and operations problem. Cutting plane methods (such as this benchmark) are inherently limited by the complexity of the weekly subproblems in parallel as well as the growing size of the master problem. 
    \item \textit{Decentralized Single Threaded (DST) Benchmark}: This benchmark is based on the current state-of-the-art decentralized methods highlighted in \cite{javad,ramanan2017asynchronous,asynch2019uc}. It employs a region based decomposition of the power network topology. An ADMM based technique is used to iteratively balance the electricity flow among the regions themselves, which iteratively leads to convergence to the optimal cost. The DST benchmark employs only a single thread at every region and as a result suffers significant performance degradation compared to DMT. 
    \item \textit{Decentralized Solver Driven Multithreading (SMT) Benchmark}: This benchmark augments DST with solver (i.e. Gurobi) driven multithreading. In this strategy, we employ the same region driven decomposition with ADMM based flow balance. However, instead of every region being given a single thread, multiple threads are assigned to the optimization solver itself at every region. Consequently, the mechanics of multithreading are completely handled by the solver without any added decomposition strategies from the formulation perspective. Therefore, such a strategy represents the most stringent, state-of-the art benchmark for comparing the decentralized multithreading scheme proposed in this paper.
\end{enumerate}
}

\subsection{Computational Results and Observations}
  {
\begin{table}[!htb]
\centering
 \captionsetup{margin={0.02\textwidth,0pt},singlelinecheck=false}
\caption{Centralized Results Objective Costs (USD $10^4$)}
\label{tab:centralized}
\begin{tabular}{c|c|c|c|c|c}
CGD & Variables & Ops & CBM & DC & Obj\\
\hline
        4 & 100800 & 641.70 & 69.80  & 2.66  & 714.16\\
        8 & 201600 & 626.66 & 69.81 & 5.12 & 701.60\\
        12 & 403200 & 624.44 & 69.84 & 7.58 & 701.86
\end{tabular}
\vspace{-2mm}
\end{table}}
\begin{table*}
\centering
         \captionsetup{margin={0.1\textwidth,0pt},singlelinecheck=false}
        \caption{Final Objective Costs (USD $10^4$) of Decentralized Joint CBM and Operations framework}
        \label{tab:cost}
        \begin{tabular}{c|cccc|cccc|cccc}
        No. Of  & \multicolumn{4}{c|}{4 CGD} & \multicolumn{4}{c|}{8 CGD} & \multicolumn{4}{c}{12 CGD}\\\cline{2-13}
        Regions & Ops& CBM & DC & Obj & Ops & CBM & DC & Obj & Ops & CBM & DC &Obj  \\ 
        \hline
        50 & 641.16&	69.82&	6.85&	717.82&			627.89&	69.80&	6.96&	704.65&			626.12&	69.86&	9.96&	705.95\\
        60 & 641.05&	69.82&	8.79&	719.66&			627.03&	69.83&	8.96&	705.82&			625.70&	69.80&	8.43&	703.94\\
        75 & 640.32&	69.84&	10.48&	720.65&			626.82&	69.87&	10.72&	707.40&			626.78&	69.81&	8.34&	704.93\\
        80 & 638.32&	69.83&	12.37&	720.52&			625.13&	69.81&	12.24&	707.18&			624.93&	69.83&	11.27&	706.03\\
        90 & 639.25&	69.84&	12.52&	721.61&			623.68&	69.82&	13.17&	706.67&			623.64&	70.01&	18.82&	712.47\\
        100 &637.93&	69.85&	14.34&	722.12&			622.80&	69.91&	18.05&	710.76&			620.61&	69.88&	20.07&	710.56\\
        \end{tabular}
\vspace{-2mm}
\end{table*}
\begin{table*}
\centering
        \captionsetup{margin={0.1\textwidth,0pt},singlelinecheck=false}
        \caption{Computational Performance of Decentralized Joint CBM and Operations framework}
        \label{tab:cp}
        \begin{tabular}{c|c|cc|cc|cc}
        No. Of  & No. Of &\multicolumn{2}{c|}{4 CGD} & \multicolumn{2}{c|}{8 CGD} & \multicolumn{2}{c}{12 CGD}\\\cline{3-8}
        Regions & Threads & Opt Gap(\%) &Time (mins) & Opt Gap(\%) & Time (mins) & Opt Gap(\%) & Time(mins)\\ 
        \hline
        50 & 600 & 0.51 & 33.77 & 0.43 & 52.39 & 0.58 &122.74\\
        60 & 720 & 0.77 & 44.58 & 0.59 & 53.54 & 0.29 &170.89\\
        75 & 900 & 0.90 & 49.8 & 0.82 & 219.17 & 0.43 &111.50\\
        80 & 960 & 0.89 & 31.49 & 0.79 & 50.98 & 0.59 &56.50\\
        90 & 1080 & 1.04 & 33.24 & 0.71 & 35.02 & 1.51 &63.06\\
        100 & 1200 & 1.11 & 41.37 & 1.2 & 39.88 & 1.23 &368.98\\
        \end{tabular}
\vspace{-2mm}
\end{table*}
\begin{table*}
\centering
        \captionsetup{labelfont={bf},margin={0.3\textwidth,0pt},singlelinecheck=false}
        \caption{ Centralized Solution Benchmarking}
        \label{tab:cp2}
         \begin{tabular}{c|c|cccccc}
        CGD & Time & \multicolumn{6}{c}{\textbf{DMT Speedup}}\\\cline{3-8}
         & (mins) & \textbf{50 Regions} & \textbf{60 Regions} & \textbf{75 Regions} & \textbf{80 Regions} & \textbf{90 Regions} & \textbf{100 Regions}\\ 
        \hline
        4 & 120.85 & \textbf{3.58} & \textbf{2.71} & \textbf{2.43} & \textbf{3.84} &	\textbf{3.64} & \textbf{2.92}\\
        8 & 611.43 & \textbf{11.67} & \textbf{11.42} & \textbf{2.79} & \textbf{11.99} & \textbf{17.46} & \textbf{15.33}\\
        12 & 2846.17 & \textbf{23.19} & \textbf{16.66} & \textbf{25.53} & \textbf{50.38} & \textbf{45.14} & \textbf{7.71}
        \end{tabular}
\vspace{-2mm}
\end{table*}
\begin{table*}[!htb]
\centering
        \captionsetup{labelfont={bf},margin={0.3\textwidth,0pt},singlelinecheck=false}
        \caption{ Two Stage APMII Benchmark}
        \label{tab:apm2}
        \setlength{\tabcolsep}{3pt}
         \begin{tabular}{c|c|c|c|ccccccc}
        CGD & Convergence & Obj	&  (\%) Gap from  & \multicolumn{6}{c}{\textbf{Lower Bound on DMT Speedup}}\\\cline{5-10}
        & Status\footnotemark & & centralized &\textbf{50 Regions} & \textbf{60 Regions} & \textbf{75 Regions} & \textbf{80 Regions} & \textbf{90 Regions} & \textbf{100 Regions}\\
        \hline
        4 & NC & 712.37 & 0.23 & \textbf{21.32} & \textbf{16.15} & \textbf{14.46} & \textbf{22.86} & \textbf{21.66} & \textbf{17.40} \\
        8 & NC & 699.93 & 0.22 &  \textbf{13.74} & \textbf{113.45} & \textbf{3.29} & \textbf{14.12} & \textbf{20.56} & \textbf{18.06}\\
        12 & NC & 699.62 & 0.26 &  \textbf{5.87} & \textbf{4.21} & \textbf{6.46} & \textbf{12.74} & \textbf{11.42} & \textbf{1.95}
        \end{tabular}
\vspace{-2mm}
\end{table*}
\begin{table*}[!htb]
\centering
         \captionsetup{labelfont={bf},margin={0.15\textwidth,0pt},singlelinecheck=false}
        \caption{ Decentralized Solver Driven Multithreading (SMT) Benchmark}
        \label{tab:gmt}
         \begin{tabular}{c|ccc|ccc|ccc}
        No. Of  & \multicolumn{3}{c|}{4 CGD} & \multicolumn{3}{c|}{8 CGD} & \multicolumn{3}{c}{12 CGD}\\\cline{2-10}
        Regions & Opt Gap (\%) & Time & \textbf{DMT} & Opt Gap (\%) & Time & \textbf{DMT} & Opt Gap (\%) & Time & \textbf{DMT} \\ 
        & & (mins) & \textbf{Speedup} & & (mins) & \textbf{Speedup} & & (mins) & \textbf{Speedup} \\
        \hline
        50  & 0.38 & 197.62	          & \textbf{5.85} & 0.31  & 306.11          & \textbf{5.84}  & 0.29  & 278.46            & \textbf{2.27} \\
        60  & 0.36 & 113.87	          & \textbf{2.55} & 0.50  & 155.77          & \textbf{2.91}  & 0.18  & 301.56            & \textbf{1.76} \\
        75  & 0.43 & 181.06	          & \textbf{3.64} &  -    & \textit{210.05} &  *             &  -    & \textit{276.25}   &  *   \\
        80  & 0.41 & 106.87	          & \textbf{3.39} &  -    & \textit{244.62} &  *             & 0.22  & 194.45            & \textbf{3.44} \\
        90  & 0.39 & 83.51	          & \textbf{2.51} &  -    & \textit{199.73} &  *             &  -    & \textit{203.27}   &  *   \\
        100	&  -   &  \textit{182.28} &   *           &  -    & \textit{228.17} &  *             &  -    & \textit{209.23}   &  *  
        \end{tabular}
\end{table*}
        
 
We present our results in terms of the calculated speedup offered by the DMT algorithm as compared to the corresponding benchmarking method. The speedup is calculated as the ratio of the solution times incurred by the benchmark with respect to those of the DMT. 
\subsubsection{Centralized Benchmark Comparison}
Table \ref{tab:centralized} depicts the operational, CBM, demand curtailment and total costs of the centralized algorithm with varying number of CGD values. We use the centralized results for benchmarking the solution quality of our decentralized, multithreaded algorithm. Table \ref{tab:cost} depicts the operational (Ops), CBM, Demand Curtailment (DC) and Total Objective cost with respect to increasing CGD values with varying number of regions. Further, Table \ref{tab:cp} shows the computational performance of the decentralized algorithm with respect to the optimality gap, the computational time incurred as well as the total number of threads used. Specifically, we show the results with varying CGD and region cases.

We observe that with increasing complexity of the problem as indicated by higher CGD values, the computational time keeps increasing. Further, Table \ref{tab:cost} shows that our decentralized approach yields highly stable solutions with respect to optimal maintenance decisions as well as operational decisions. On the other hand Table \ref{tab:cp} yields numerous interesting insights concerning the computational performance of the decentralized algorithm. First, we observe that with increasing number of regions, the time incurred by our algorithm does not increase by a significant amount. In fact, our approach is seen to be highly stable with increasing number of regions thereby highlighting scalability. Moreover, we can also note that with increasing complexity of the problem denoted by the CGD value, the time incurred increases approximately in a linear fashion thereby demonstrating computational efficiency. Lastly, we observe that in terms of quality of solution as indicated by the optimality gap, our algorithm retains its stability with increasing number of regions across varying CGD values. 

Table \ref{tab:cp2} presents the performance in terms of the speedup offered by DMT in comparison with the centralized method. We observe that the DMT algorithm consistently solves the joint problem at least three times as fast as the centralized counterpart. In fact, it can be seen from Table \ref{tab:cp2}, that the DMT can be 50 times faster than the conventional centralized solution method. 

\subsubsection{Two Stage APMII Benchmark Comparison}
Table \ref{tab:apm2} represents the results relating the Two Stage APMII benchmark with respect to varying CGD. For these benchmark experiments, we provided a maximum time limit of 12 hours. The weekly operational subproblems in our benchmark implementation were also parallelized using MPI. We note that none of the experiments pertaining to APMII converged within the specified time limit. Therefore, we present only the projected speedup values, which evidently represents the lower bound on the speedup offered by the DMT algorithm. 

Table \ref{tab:apm2} clearly demonstrates that the DMT can provide a speedup by a factor of 22 in certain cases. The speedup values largely remain stable across increasing number of regions however, they decrease slightly with increasing CGD complexity. Even at its slowest, DMT is still approximately at least twice as fast as the APMII method. Therefore, results in Table \ref{tab:apm2} indicate that the DMT algorithm outperforms state-of-the-art Two Stage APMII method in all the cases that we have studied.

\subsubsection{Decentralized Single Threaded (DST) Benchmark}
\begin{table}[!htb]
\centering
         \captionsetup{labelfont={bf},margin={0.02\textwidth,0pt},singlelinecheck=false}
        \caption{ DST Benchmark with respect to CGD=4}
        \label{tab:dst1}
         \begin{tabular}{c|ccccccc}
        Regions & Opt Gap (\%) & Time(mins) & \textbf{DMT Speedup} \\ 
        \hline
        50 & 0.33 & 594.32 & \textbf{17.60}\\
        60 & 0.30 & 943.80 & \textbf{21.17}\\
        75 & 0.46 & 391.69 & \textbf{7.87}\\
        80 & 0.40 & 260.46 & \textbf{8.27}\\
        90 & 0.37 & 197.03 & \textbf{5.93}\\
        100 & 0.50 & 107.61 & \textbf{2.60}
        \end{tabular}
\vspace{-5mm}
\end{table}
\begin{table}[!htb]
\centering
         \captionsetup{labelfont={bf},margin={0.02\textwidth,0pt},singlelinecheck=false}
        \caption{ Performance Comparison of DST with 100 Region Decomposition}
        \label{tab:dst2}
         \begin{tabular}{c|ccccccc}
        CGD & Opt Gap (\%) & Time(mins) & \textbf{DMT Speedup} \\ 
        \hline
        4 & 0.50 & 107.61 & \textbf{2.60}\\
        8 & 0.49 & 1,967.09 & \textbf{49.33}\\
        12 & 0.39 & 2,776.94 & \textbf{7.53}\\
        \end{tabular}
\end{table}
For comparisons with respect to DST, convergence was very slow on account of no multithreading support available to the solver at the regional level. We restricted our benchmark experiments to the 4 CGD case across all regional decompositions and the 100 region case with varying computational complexity depicted in Tables \ref{tab:dst1}, \ref{tab:dst2} respectively.

Table \ref{tab:dst1} demonstrates that the DMT algorithm can be faster by up to a factor of 18 with the 4 CGD case. While the speedup of DMT decreases with increasing number of regions, it is still faster than DST at least by a factor of 2.6 for the 100 region case. \footnotetext{  {A convergence status of NC refers to Non Convergence}}On the other hand, analysis of results in Table \ref{tab:dst2} reveal that DMT can be 50 times faster than DST while still delivering a healthy speedup across all degrees of complexity. Tables \ref{tab:dst1} and \ref{tab:dst2} show the degradation of performance with increasing complexity as well as regions as evidenced by convergence times stretching up to 48 hours (for the 12 CGD, 100 regions case).

\subsubsection{Decentralized Solver Driven Multithreading Benchmark} 
Table \ref{tab:gmt} represents the results obtained by allowing the solver to leverage multithreading itself. In this case, in order to compare with the DMT technique, we allocate 12 threads to the solver per region while keeping the same experimental settings. Table \ref{tab:gmt} depicts  the optimality gap, the time as well as the speedup offered by DMT in comparison with the benchmark method. Rows marked * represent instances where the solver ran out of memory along with the time taken to reach the out of memory exception.

Table \ref{tab:gmt} reveals several interesting trends. First, we can see that even with a state-of-the-art solver like Gurobi, the DMT method delivers consistent speedup values of up to a factor of 6. Further, out of memory exceptions increase with the rise in complexity. This can be directly attributed to the increasing number of binary variables encountered with increasing CGD values. As a result, more memory is consumed by the multithreading schemes on commercial solvers, leading to a poor space complexity and scalability.
\color{black}
\section{Conclusion and Future Work}\label{sec:mconclusion}
In this paper we develop a decentralized, multithreaded framework for the joint CBM and operational problem designed for large scale power systems.  Our solution involves, decomposing a given power network topology into multiple regions and using ADMM to formulate a joint optimization model. Such a formulation based on decentralization, allows preservation of regional data privacy. We solve the optimization model in a decentralized manner wherein each region holds its own local subproblem and cooperates with its neighbors by exchanging flow estimates. We further leverage multithreading at every regional subproblem to bolster the computational efficiency of our solution.   

We demonstrate the convergence of our algorithm based on experiments on the large scale IEEE 3012 bus case incorporating varying degrees of region decompositions. We benchmark our methodology against four different computational paradigms i.e., centralized, two stage cutting plane, decentralized single threaded as well as decentralized solver driven multithreaded methods. Our results show that our decentralized multithreaded algorithm consistently outperforms all four benchmarking methods offering as high as 50x speedups with a stable solution quality as well. The results demonstrate that our decentralized algorithm can provide good solution quality, scalability and efficiency with full privacy of sensor data while being robust to varying problem complexities. As part of our future  work,  we  intend  to  explore  applications of the proposed framework { into coordination mechanisms for different participants within deregulated electricity markets,} and computational algorithms that can harness further efficiency through  asynchronous  model  of  operation  as  well as integrating differential privacy for consensus.


\section*{Acknowledgement}
This paper is dedicated to Sudha Venkat Ramanan, a constant source of inspiration and whose love, support and understanding played a major role in making this research happen.
\bibliography{artifact}
\bibliographystyle{ieeetr}
 
\section*{Appendix: Implementation Notes}
The experiments were performed on the Hive supercomputer at the Georgia Institute of Technology, Atlanta, GA, USA. The codebase used in the experiments consisted of C++ as well as Python based optimization modules that we developed for Gurobi 7.5. Python 2.7 was used for implementation including the creation of the joint CBM and operations optimization models. 
To improve computational efficiency, the solver modules were written in C++ with Cython wrappers for easy and fast access from Python.

There were numerous reasons for adopting a mixed C++ and Python oriented framework. Support for multithreading frameworks like OpenMP is largely absent in Python. For solving the same optimization model, there was a 10x reduction in compute time in case of a C++ implementation as compared to a Python based one.

\vskip 0pt plus -1fil
\begin{IEEEbiographynophoto}{Paritosh Ramanan}
Paritosh Ramanan is a Postdoctoral Fellow with the Georgia Institute of Technology in Atlanta, Georgia. He got his PhD in Computational Science and Engineering from the H. Milton Stewart School of Industrial and Systems Engineering at Georgia Institute of Technology in Atlanta, Georgia in 2020. Prior to his PhD he earned a Masters in Computer Science from Georgia State University in Atlanta, Georgia in 2015 and obtained his Bachelors in Information Systems from Birla Institute of Technology and Science (BITS) Pilani, Goa Campus in 2013. His research focuses on developing decentralized algorithms for improved computational performance of large scale optimization problems through the use of parallel and distributed computing paradigms. 
\end{IEEEbiographynophoto}
\vskip 0pt plus -1fil
\begin{IEEEbiographynophoto}{Murat Yildirim}
Dr. Murat Yildirim is an Assistant Professor in the Department of Industrial and Systems Engineering at Wayne State University. Prior to joining Wayne State, he worked as a postdoctoral fellow at the Georgia Institute of Technology (2016-2018), and obtained a Ph.D. degree in Industrial Engineering, and B.Sc. degrees in Electrical and Industrial Engineering from the same institution. Dr. Yildirim's research interest lies in advancing the integration of mathematical programming and data analytics in large scale energy systems. Specifically, he focuses on the modeling and the computational challenges arising from the integration of real-time sensor inferences into large-scale mixed integer programs (MIPs) used for optimizing and controlling networked systems.
\end{IEEEbiographynophoto}
\vskip 0pt plus -1fil
\begin{IEEEbiographynophoto}{Nagi Gebraeel}
Dr. Nagi Gebraeel is the Georgia Power Early Career Professor and Professor in the Stewart School of Industrial and Systems Engineering at Georgia Tech. His research interests lie at the intersection of industrial predictive analytics and decision optimization models for large scale power generation applications. Dr. Gebraeel serves as an associate director at Georgia Tech's Strategic Energy Institute and the director of the Analytics and Prognostics Systems laboratory at Georgia Tech's Manufacturing Institute. Dr. Gebraeel was the former president of the Institute of Industrial Engineers (IIE) Quality and Reliability Engineering Division, and is currently a member of the Institute for Operations Research and the Management Sciences (INFORMS).
\end{IEEEbiographynophoto}
\vskip 0pt plus -1fil
\begin{IEEEbiographynophoto}{Edmond Chow}
Edmond Chow is an Associate Professor in the School of Computational Science and Engineering at Georgia Institute of Technology.  He previously held positions at D. E. Shaw Research and Lawrence Livermore National Laboratory. His research is in developing numerical methods specialized for high-performance computers and applying these methods to enable the solution of large-scale physical simulation problems in science and engineering. Dr. Chow received the 2009 ACM Gordon Bell prize and a 2002 Presidential Early Career Award for Scientists and Engineers (PECASE). He is a Fellow of the Society for Industrial and Applied Mathematics.
\end{IEEEbiographynophoto}
\end{document}